\documentstyle[11pt,aaspp4,flushrt,tighten,psfig]{article}  

\begin{document}

\title{New Results from the X--ray and Optical\footnote{Based on
observations performed at the European Southern Observatory, Paranal,
Chile} Survey of the Chandra Deep Field South: The 300ks Exposure}

\author{P.Tozzi\altaffilmark{1}, P. Rosati\altaffilmark{2},
M. Nonino\altaffilmark{1}, J. Bergeron\altaffilmark{2},
S. Borgani\altaffilmark{3}, R. Gilli\altaffilmark{4},
R. Gilmozzi\altaffilmark{2}, N. Grogin\altaffilmark{5},
G. Hasinger\altaffilmark{6}, L. Kewley\altaffilmark{4},
A. Koekemoer\altaffilmark{5}, C. Norman\altaffilmark{4,5},
E. Schreier\altaffilmark{5}, G. Szokoly\altaffilmark{6},
J.X. Wang\altaffilmark{4,7}, W. Zheng\altaffilmark{4}, A. Zirm
\altaffilmark{4}, and R. Giacconi\altaffilmark{4,8} }

\affil{$^1$Osservatorio Astronomico di Trieste, via G.B. Tiepolo 11,
I--34131, Trieste, Italy} 
\affil{$^2$European Southern Observatory,
Karl-Schwarzschild-Strasse 2, D-85748 Garching, Germany}
\affil{$^3$INFN, c/o
Dip. di Astronomia dell'Universit\`a, via Tiepolo 11, I--34131,
Trieste, Italy} 
\affil{$^4$Dept. of Physics and Astronomy, The Johns Hopkins
University, Baltimore, MD 21218, USA} 
\affil{$^5$ Space Telescope Science Institute, 3700
S. Martin Drive, Baltimore, MD 21210, USA} 
\affil{$^6$Astrophysikalisches Institut, An der Sternwarte 16, Potsdam
14482 Germany} 
\affil{$^7$ Center for Astrophysics, University of Science and Technology of China,
Hefei, Anhui, 230026, P. R. China}
\affil{$^8$ Associated
Universities Inc., 1400, 16th st. NW, Washington DC 20036, USA}

\begin{abstract}

We present results from 300 ks of X--ray observations of the Chandra
Deep Field South.  The field of the four combined exposures is now
0.1035 deg$^2$ and we reach a flux limit of $10^{-16}$ erg s$^{-1}$
cm$^{-2}$ in the 0.5--2 keV soft band and $10^{-15}$ erg s$^{-1}$
cm$^{-2}$ in the 2--10 keV hard band, thus a factor 2 fainter than the
previous 120 ks exposure.  The total catalogue is composed of 197
sources including 22 sources detected only in the hard band, 51 only
in the soft band, and 124 detected in both bands.  We have now the
optical spectra for 86 optical counterparts.

The LogN--LogS relationship of the whole sample confirms the
flattening with respect to the ASCA hard counts and the ROSAT soft
counts.  The average logarithmic slope of the number counts is $\alpha
= 0.66\pm 0.06$ and $\alpha = 0.92\pm 0.12$ in the soft and hard band
respectively.  Double power--law fits to the differential counts show
evidence of further flattening at the very faint end to a slope of
$0.5\pm 0.1$ and $0.6\pm 0.2$ in the soft and in the hard band
respectively.  We compute the total contribution to the X--ray
background in the 2--10 keV band, which now amounts to $(1.45\pm 0.15)
\times 10^{-11}$ erg cm$^{-2}$ s$^{-1}$ deg$^{-2}$ (after the
inclusion of the ASCA sources to account for the bright end) to a flux
limit of $10^{-15}$ erg s$^{-1}$ cm$^{-2}$.  This corresponds to
60-90\% of the unresolved hard X--ray background (XRB), given the
uncertainties on its actual value.

We confirm previous findings on the average spectrum of the sources
which is well described by a power law with $\Gamma = 1.44 \pm 0.03$,
and the progressive hardening of the sources at lower fluxes.  In
particular we find that the average spectral slope of the sources is
$\Gamma = 1.26 \pm 0.04$, flatter than the average, for fluxes lower
than $9 \times 10^{-15}$ erg s$^{-1}$ cm$^{-2}$ in the hard band.  The
hardening of the spectra is consistent with an increasing fraction of
absorbed objects ($N_H > 10^{22}$ cm$^{-2}$) at low fluxes.

From 86 redshifts available at present, we find that hard sources
have on average lower redshifts ($z\leq 1$) than soft sources.
Their typical luminosities and optical spectra show that
most of these sources are obscured AGNs, as expected by AGN population
synthesis models of the X--ray background.  We are still in the
process of finding hard sources that constitute the remaining fraction
of the total XRB.

Most of the sources detected only in the soft band appear to be
optically normal galaxies with luminosities $L_X\simeq 10^{40}-
10^{42}$ erg s$^{-1}$.  This population appears to be a mix of
normal galaxies, possibly with enhanced star formation, and galaxies
with low--level nuclear activity.

\end{abstract}


\newpage

\section{INTRODUCTION}

We present here the analysis of the 300 ks exposure of the Chandra
Deep Field South (Giacconi et al. 2000; hereafter Paper I) which
reaches an on--axis flux limit of $S=10^{-15}$ erg s$^{-1}$ cm$^{-2}$
and $S=10^{-16}$ erg s$^{-1}$ cm$^{-2}$ in the hard (2--10 keV) and
soft (0.5--2 keV) bands respectively.  This is a factor of two deeper
than the exposure presented in Paper I, and significantly deeper than
the Chandra field presented by Mushotzky et al. (2000) and the XMM
observations of the Lockman Hole by Hasinger et al. (2001a).  With
this dataset we have a similarly large survey area and limiting flux
to the Chandra Deep Field North (Hornschemeier et al. 2000; 2001;
Brandt et al. 2001).  Results from the total exposure of one million
seconds will be reported elsewhere (Rosati et al. 2001).

With this new data we confirm several results that were derived in the
initial 120 ks dataset, and extend our analysis by resolving a larger
fraction of the soft and hard X--ray background. The larger number of
source photons allows a more accurate determination of the average
spectral properties of the sample. The general properties of the X-ray
background are now well established. This paper focuses on the nature
of the sources that make up the background, especially in the hard
band, taking advantage of new measured redshifts. We currently have 86
spectroscopic identifications from the on--going observational
campaign with the VLT.  Combining the X--ray and optical data, we can
now study the intrinsic luminosities for a significant subset of
sources, thus allowing for a characterization of the source
population.

The paper is structured as follows.  In \S 2 we describe the data
acquisition and analysis.  In \S 3 we present the results of the
analysis of the X--ray data.  In \S 4 we discuss some of the
characteristics of the observed sources using the optical information
available, and discuss the possible implications of our findings. Our
conclusions are summarized in \S 5.

\section{OBSERVATIONS AND DATA REDUCTION}

\subsection{X--Ray data}

The Chandra Deep Field South (CDFS) data presented here have been
obtained by adding two exposures to the former 120 ks exposure
described in Paper I.  The two new exposures were taken with the
Advanced CCD Imaging Spectrometer--Imaging (ACIS--I) detector at a
temperature of -120 C.  The first observation (Obs ID 441) was taken
on 2000-5-27, in the faint mode, for a total 60 ks exposure. The
second observation (Obs ID 582) was taken on 2000-6-3/4 in the faint
mode, for a total 120 ks exposure.

The data were reduced using the same techniques already described in
Paper I (CIAO software release V1.5, see http://asc.harvard.edu/cda).
The most important difference is the lower temperature of the detector
(-120 C) that requires a new set of quantum efficiency uniformity
files.  At the temperature of -120 C the charge transfer inefficiency
is largely reduced, and the effective area is greatly improved,
especially in the hard band.

The data were filtered to include only the standard event grades
0,2,3,4 and 6.  All hot pixels and columns were removed.  We removed
the flickering pixels with more than two events contiguous in time,
where a single time interval was set to 3.3 s.  Time intervals with
background rates larger than $3$ $\sigma$ over the quiescent value
($0.31$ counts s$^{-1}$ per chip in the 0.3--10 keV band) were
removed. The 4 observations give a total exposure time of 303 ks.  The
four observations had different roll angles, enabling a total coverage
of 0.1035 deg$^2$.  The exposure time across the field varies from a
maximum of 303 ks in the center to a minimum of 25 ks in the corners.
The area covered by 303 ks of exposure is 0.0636 deg$^2$.  The
resulting exposure map is the sum of the exposure maps of the single
observations weighted by the corresponding exposure times.

The detection strategy has been modified with respect to Paper I for a
faster detection algorithm, allowing a large number of simulations to
be carried out.  We used Sextractor (Bertin \& Arnouts 1996) on the
0.5--7 keV image rebinned by a factor of two, so that one image pixel
corresponds to $0.984\arcsec$.  Sextractor detection parameters were
chosen as a result of simulations and we adopted a detection threshold
of 2.4, with a Gaussian filter with 1.5 arcsec FWHM and a minimum area
of 5 pixels.  Sextractor is not tailored for use with a very low and
sparse background as in ACIS--I. We used a modified version of the
program to allow an external map to be used as local background. The
smoothed map of the background was computed from the data itself after
the removal of the sources down to a very low threshold.  This
modified detection algorithm is several order of magnitude faster than
the wavelet detector algorithm of Rosati et al. (1995) or WAVDETECT in
the CIAO software (Freeman et al. 2001).

\begin{figure}
\centerline{\psfig{figure=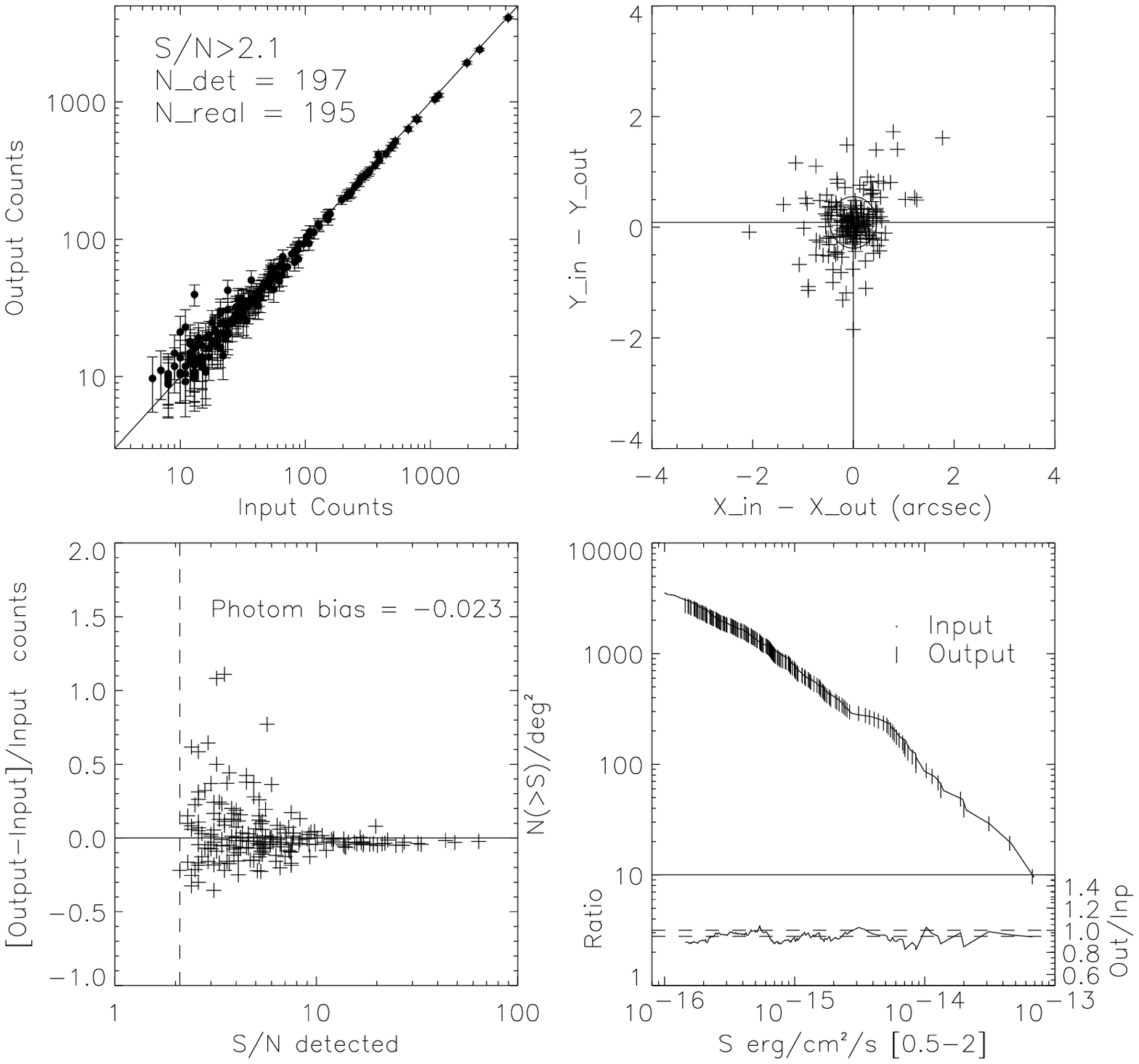,height=6.in,width=6.in}}
\caption{Analysis of one of the simulations performed with MARX for
the 300ks exposure.  We show the measured counts vs input counts for
each source (panel a); the difference between the input source
positions and the measured positions (panel b); the relative  errors
on the detected counts as a function of the S/N ratio (panel c); the
output vs the input number counts (panel d).  This analysis refers to
the soft band.
\label{fig1}}
\end{figure}

The MARX v3.0 simulator (Wise et al. 2000) was used to generate the
photon distribution of sources drawn from an input LogN--LogS modeled
on the observed one.  Sources were placed randomly on the ACIS
detector, and each one of them has been simulated as a point source as
defined in MARX.  Four different observations with the same roll
angles and exposure maps as the real observations are simulated for a
given distribution of sources.  The final event file is obtained by
merging four event files.  The sky background was created as a poisson
realization of the observed background, which was obtained by removing
all the detected sources from the real 300ks image.  The same
detection procedure was carried out with 100 simulations.  We stress
that each of these simulations includes all the details of a realistic
observation, including the different orientation of the asymmetric
PSF, and the variations in the background due to the edges of the
chips and to the removed columns.  The only ingredient missing in the
simulations are the presence of extended sources and the clustering of
the point sources.  Both have a negligible impact on the efficiency of
our detection procedure.  We run our detection algorithm on the
simulated fields and assessed the photometric and astrometric
accuracy, completeness and spurious source contamination, as well as
the sky coverage model by comparing the input and output LogN--LogS
shown in Figure \ref{fig1}.  Simulations show that the detection
efficiency of Sextractor is comparable with that of the wavelet
algorithm.  Moreover, our $S/N$ criterium allows a direct and simple
definition of the sky coverage (see section \S 3.2) to give accurate
number counts.

We measure the signal to noise ratio of all the candidate detections
in the area of extraction of each source, which is defined as a circle
of radius $R_s=2.4\times FWHM$ (with a minimum of 5 pixels of radius).
The FWHM is modeled as a function of the off-axis angle to reproduce
the broadening of the PSF. In each band a detected source has a $S/N
\equiv S/\sqrt{S+2 B}> 2.1$ within the extraction area of the image.
Here $B$ are the background counts found in an annulus with outer
radius $R_S+12''$ and an inner radius of $R_S+2''$, after masking out
other sources, rescaled to the extraction region.  A combined catalog
is then produced matching the two.  We stress that the condition of
having $S/N > 2.1$ in the extraction area corresponds to a high
significance detection (the faintest detected sources have more than
10 counts).  Here we used a threshold higher than the one used in
Paper I ($S/N > 2.0$) to reduce the number of spurious sources.  Our
catalog now typically includes less than 5 spurious sources, as tested
with simulations.

Source counts are measured with simple aperture photometry within
$R_s$ in the soft and hard bands separately.  Simulations have shown
that such aperture photometry leads to an underestimate of the source
count rate by approximately $-4$\% (see panel c of Figure \ref{fig1}).
We will correct such photometric bias before converting count--rates
in energy flux.

The count--rate to flux conversion factors in the 0.5--2 keV and in
the 2--10 keV bands were computed using the response matrices.  We
quote the fluxes in the canonical 2--10 keV band, as extrapolated from
counts measured from the 2--7 keV band, in order to have a direct
comparison with the previous results.  The conversion factors are
$(4.6\pm 0.1) \times 10^{-12}$ erg s$^{-1}$ cm$^{-2}$ per count
s$^{-1}$ in the soft band, and $(2.9\pm 0.3) \times 10^{-11}$ erg
s$^{-1}$ cm$^{-2}$ per count $\rm s^{-1}$ in the hard band assuming a
Galactic absorbing column of $8 \times 10^{19}$ cm$^{-2}$ and a photon
index $\Gamma = 1.4$.  The uncertainties in the conversion factors
reflect the range of possible values for the effective photon index
shown by the sources in our sample: $\Gamma = 1.1 - 1.7$.

The conversion factors were computed at the aimpoint.  Before
computing the energy flux, the count rates must be corrected for
vignetting and are converted to the count rates that would have been
measured if the source were in the aimpoint.  The correction is simply
given by the ratio of the value of the exposure map at the aimpoint to
the value of the exposure map at the source position.  This is done
separately for the soft and the hard band, using the exposure maps
computed for energies of 1.5 keV and 4.5 keV.  This procedure also
accounts for the large variation in exposure time across the field of
view.

\subsection{OPTICAL AND IR DATA}

The optical counterparts of X--ray sources have been identified on
deep R band images obtained with the Focal Reducer/Low Dispersion
Spectrograph (FORS1, Appenzeller et al. 1998) at VLT--ANTU. Most of
the area covered by the 4 chips of ACIS--I has been observed in R band
with an exposure time ranging from 1.5 to 2 hours, in Service
Observing mode on October 1999--March 2000.  Shallower exposures
(about 20 minutes) have also been obtained for the outer regions
covered by the ACIS--I chips at different roll angles.  These images
were reduced in the standard way (prescan, bias, sky-flat and supersky
correction), astrometrically and photometrically calibrated and then
coadded using drizzling procedures.  The depth of the coadded image is
$R \approx 26.0 $ for the shallow part and $R \approx 26.8 $ for the
deeper part of the coadded image (at $2\sigma$ level within $2\arcsec$
aperture). The PSF in the final images is $0.6-0.8\arcsec$.  The area
covered by the deep R-band images has also been observed in I band,
with 3 hours of exposure and a limiting magnitude of $I\approx 26.5$,
with a PSF in the final image of $\approx 0.6\arcsec$.

For those sources which do not fall in the area covered by FORS1
images, the optical identification has been performed on a B band
image of 6.5 hours integration obtained during the commissioning phase
of the Wide Field Imager (Baade et al. 1999) at the 2.2 MPA--ESO
Telescope in January 1999.  The limiting magnitude is $B\simeq 26.5$
($5 \sigma)$ with a PSF of $1\arcsec$.  We have also used updated data
in J and K bands as part of the ESO Imaging Survey on the ESO NTT (see
Arnouts et al. 2001; Vandame et al. 2001).

The optical spectroscopy of X--ray counterparts with $R \geq 26$ is
underway with FORS1 at VLT--ANTU.  So far, the optical spectra were
obtained observing with the multislit mode on FORS--1 at VLT--ANTU on
October 27--30, 2000 and November 23--25, 2000.  To date, we have
obtained good S/N spectra for 86 sources, selected from a catalog
which was constructed from the 300ks image.  These redshifts are used
in the following analysis.  The list of redshifts will be presented
with the complete catalog from the 1Ms data in Giacconi et
al. (2001, in preparation), and the detailed description of the
optical spectroscopy will be presented in a future paper (Hasinger et
al. 2001b).

\section  {RESULTS}

\subsection{X--ray Spectra and Hardness Ratio}

We have performed a detailed analysis of the X--ray spectral
properties.  We have measured the average stacked spectrum of the 197
sources of the total sample.  The corresponding background was
constructed using the stacked spectrum of all the background regions
extracted around each source.  The resulting photon files were scaled
by the ratio of the total area of extraction of the sources and the
corresponding area used for the background.  Such a procedure
guarantees a correct background subtraction despite the
non--uniformities of the instrumental background across ACIS--I.  The
ancillary response matrix for the stacked spectra was obtained from
the counts--weighted average of the matrices from each of the single
sources.  Each ancillary response matrix is composed of the sum of the
matrices corresponding to that source in each single exposure,
weighted by the corresponding exposure time.  The same is done for the
response matrices.  In this way, we keep track in the most detailed
way of the characteristics of the different regions and the different
detector temperatures at the time of the observations.

\begin{figure}
\centerline{\psfig{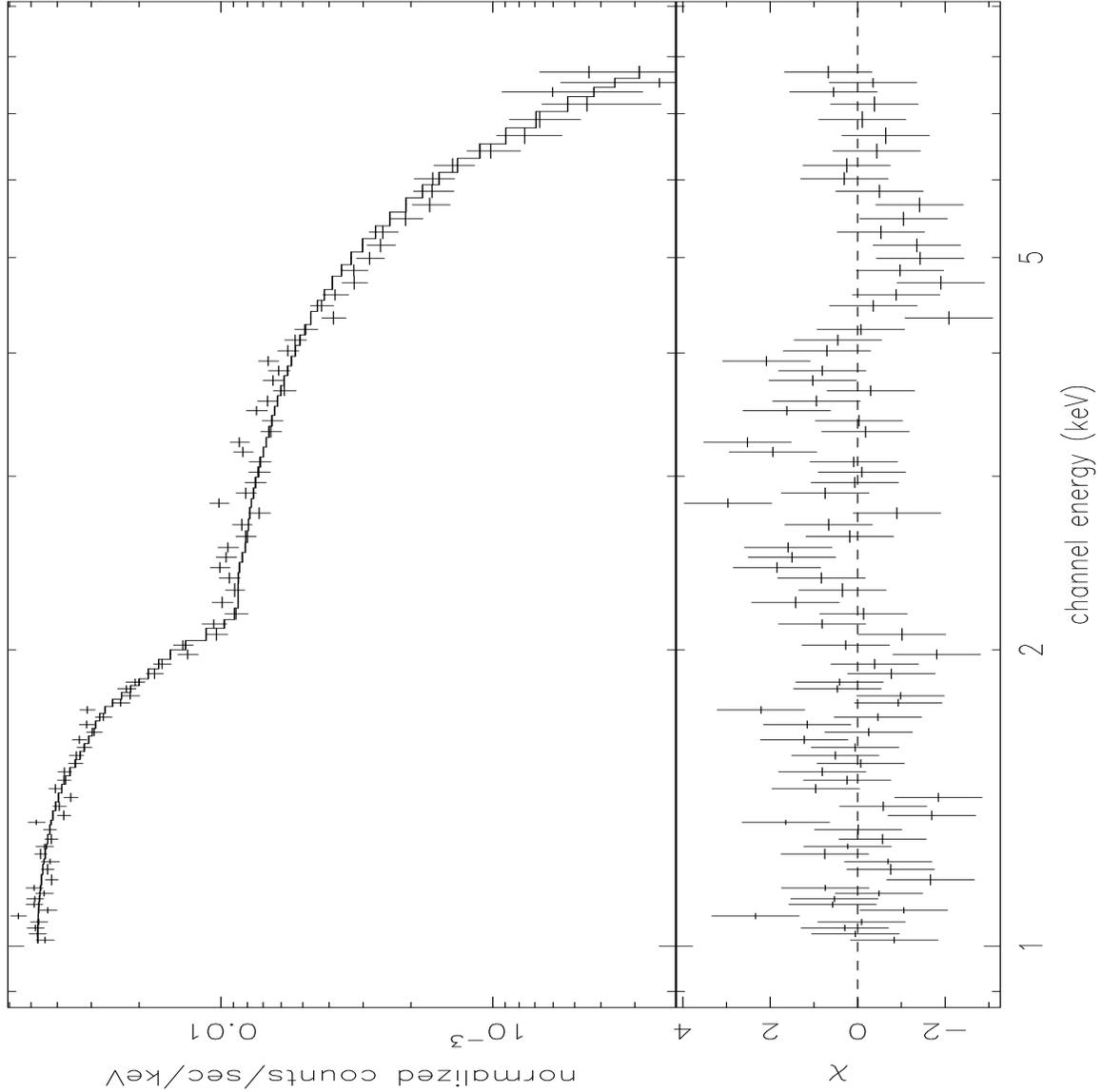}}
\caption{Stacked spectrum of the total sample of sources in the CDFS,
fitted in the energy band $1-8$ keV with an absorbed power law with
column density fixed to the Galactic value $N_H = 8\times 10^{19}$
cm$^{-2}$.  The best fit slope is $\Gamma = 1.44 \pm 0.03$ (errors at
90\% c.l.) for a $\chi_\nu^2 = 1.16$, in agreement with the average
slope of the unresolved hard X--ray background.  The solid line is the
best--fit model, while the lower panel shows the standard deviations
in each energy bin.
\label{fig2}}
\end{figure}

We used XSPEC v11.0 to compute the slope of a power law spectrum with
Galactic absorption, in the energy range 1.0--10 keV.  We exclude
energy bins below $1.0$ keV because the calibration is still uncertain
below this energy.  Such uncertainty has a small effect on the average
conversion factor, but can give wrong results on detailed fits.  For a
column density fixed to the Galactic value $N_H = 8 \times 10^{19}$
cm$^{-2}$, we obtained a photon index $\Gamma = 1.44 \pm 0.03$ (errors
refer to the 90\% confidence level) with $\chi^2_{\nu}=1.16$.  The
results of the spectral fit to the total sample of sources are shown
in Figure \ref{fig2}.  The average spectrum of the detected sources is
consistent with the measured shape of the unresolved hard background
$\langle \Gamma \rangle \simeq 1.4$, confirming previous findings with
Chandra (Paper I; Mushotzky et al. 2000).

\begin{figure}
\centerline{\psfig{figure=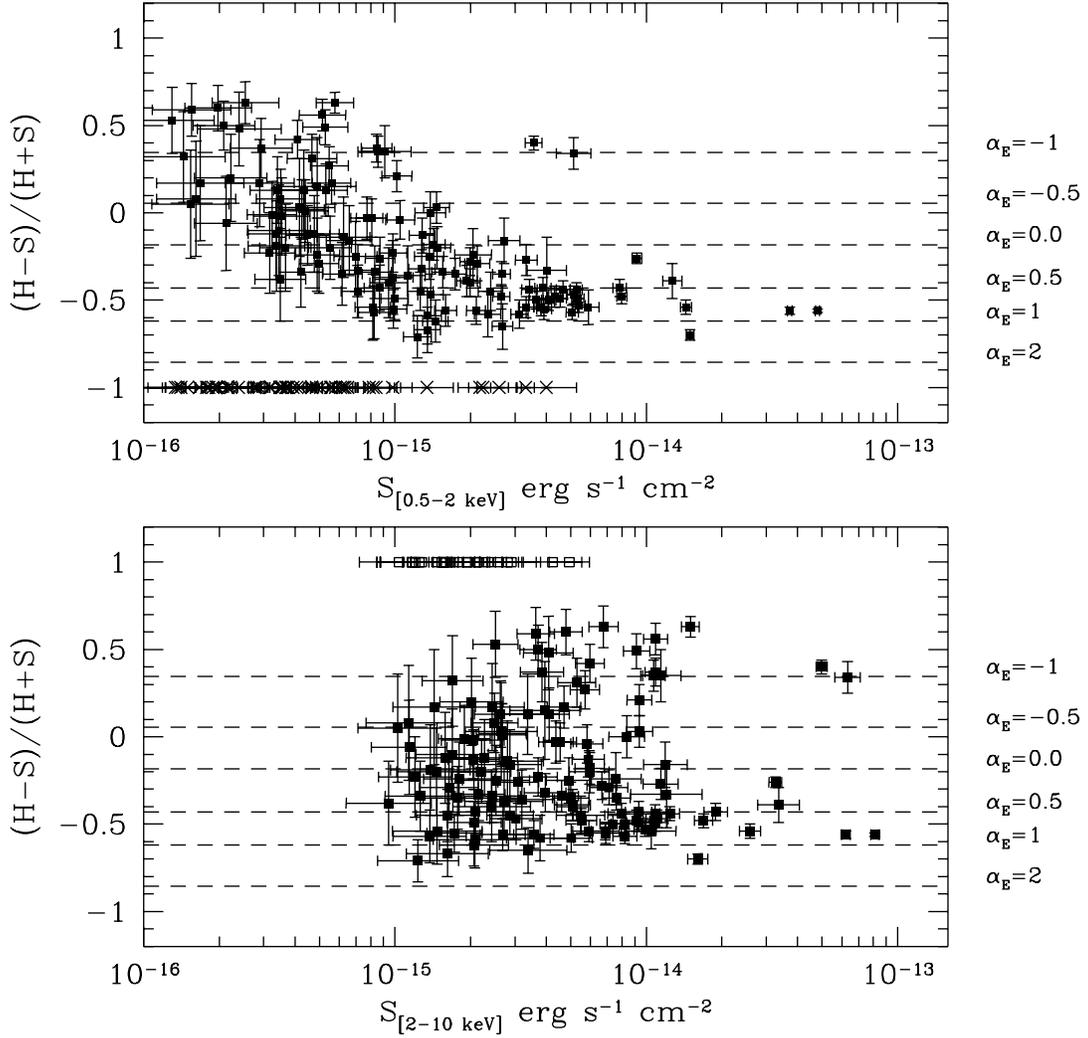,height=6.in}}
\caption{Hardness ratio for the sources detected in the soft and hard
images as a function of the flux in the corresponding band.  Sources
detected only in the hard band are shown with open squares at HR$=1$,
and the sources detected only in the soft band are shown with crosses
at HR$=-1$.  Dashed lines are power--law models with different energy
index ($\alpha_E \equiv \Gamma -1$) computed assuming the Galactic
value $N_H = 8\times 10^{19}$ cm$^{-2}$ and convoluted with a mean
ACIS--I response matrix at the aimpoint.  Note that the number of hard
sources increases at lower fluxes in both cases.
\label{fig3}}
\end{figure}

A more detailed view of the spectral properties as a function of the
fluxes comes from the hardness ratio $HR = (H-S)/(H+S)$ where $H$ and
$S$ are the net count rates in the hard (2--7 keV) and the soft band
(0.5--2 keV), respectively.  The distributions of the hardness ratios
as a function of the soft and hard flux are shown in Figure
\ref{fig3}.  In both bands, the number of hard sources increases at
lower fluxes.  This is particularly significant if we plot the
hardness ratio as a function of the hard fluxes.  The average $HR$
ranges from $HR\simeq -0.2$ for $S> 5 \times 10^{-15}$ erg s$^{-1}$
cm$^{-2}$, to $HR\simeq 0.1$ for $S< 5 \times 10^{-15}$ erg s$^{-1}$
cm$^{-2}$.  Such hardness ratio corresponds to an energy spectral
index $\alpha_E = 0$ and $\alpha_E=-0.5$ respectively.  Even if we
have already resolved the majority of the 2--10 keV background, the
remaining fraction is made by a population of sources with
progressively harder spectra.

There are 22 sources ($\simeq 11$\% of the total sample) which are
observed only in the hard band plotted at $HR=1$, and 51 sources
($\simeq $ 26\% of the total sample) which are observed only in the
soft band plotted at $HR=-1$.  Studies from ASCA suggest that each
source with nuclear activity is emitting in both bands at some level:
the soft emission is $\ge 1$\% of the hard emission for the hardest
sources (Della Ceca et al. 1999a).

In the framework of the unification model for the AGNs, we expect to
detect most of these sources in the soft and hard band if we push the
detection limit down to low fluxes.  We study the stacked spectra of
the only--soft and the only--hard subsamples, to determine the flux
level at which the single sources can be detected in the hard and soft
bands respectively.  For the sources detected only in the soft, the
stacked spectrum has an hardness ratio of $HR \geq -0.67\pm 0.05$ and
is well fitted by a power law with $\Gamma = 2.2\pm 0.2$ for a
$\chi_\nu^2 \simeq 1.2$.  This is consistent with the typical spectrum
of unabsorbed TypeI AGN, similar to the sources found at very high
fluxes by ASCA.

For the source detected only in the hard band, the hardness ratio of
the cumulative spectrum is $HR \simeq 0.64\pm 0.09$.  The fit with an
unabsorbed power law gives $\Gamma = 0.0 \pm 0.2$, but with $\chi_\nu^2
\simeq 2$.  This indicates that a single power law is a bad fit and
therefore intrinsic absorption is a necessary ingredient to describe
the spectra of the faint, hardest sources.  In fact, if we leave the
local absorption free, we find $N_H = (2.5 \pm 0.5)\times 10^{22}$
cm$^{-2}$, which is more than two orders of magnitude larger than the
Galactic value, and a power law of $\Gamma = 1.55\pm 0.3$ (errors
correspond to the 90\% confidence level).  Such level of absorption
correspond to $N_H > 10^{23}$ cm$^{-2}$ for $z>1$.  This results shows
that the progessive hardening of the sources at faint fluxes is due to
intrinsic absorption.

Moreover, the average hardness ratios derived for the two subsamples
tell us that the average spectrum of the sources undetected in one of
the two bands is similar to that of the other sources detected in both
bands at low fluxes.  In other words, we expect to detect most of them
in a longer exposure.  The average number of soft and hard counts for
the sources detected in the hard or in the soft band only, is 3.8 and
4.1 respectively (corresponding to a count rate of $\simeq 1.3 \times
10^{-5}$ cts/s).

The flux where the whole sample is detected in both the soft and the
hard band will depend on the nature of the sources themselves.  For
example, the absorption detected in the only--hard subsample is
already well in excess of $N_H = 10^{22}$ cm$^{-2}$.  If the intrinsic
absorption is rapidly increasing with decreasing flux, as predicted by
some AGN synthesis models (see Gilli, Salvati \& Hasinger 2001), then
the expected count rate of most of these hard sources can be much
lower than the expected average $\simeq 1.3 \times 10^{-5}$ cts/s and
they could be undetectable in the soft band down to fluxes as low as
$S\simeq 10^{-19}$ erg s$^{-1}$ cm$^{-2}$ (see discussion in \S 5).

In conclusion, at our present flux limits, we are still seeing
significant differences in the soft and hard subsamples.  These can
result from an admixture of different classes of objects.  This
finding encourages us to push the flux detection threshold to the
point where all the sources are detected in both bands. The value of
such a flux can be much lower than expected depending on the intrinsic
properties of the low--flux sources.  This will be investigated with
the total ($\sim$1Ms) exposure of the CDFS.

\subsection{LogN--LogS and Total Flux from Discrete Sources}

\begin{figure}
\centerline{\psfig{figure=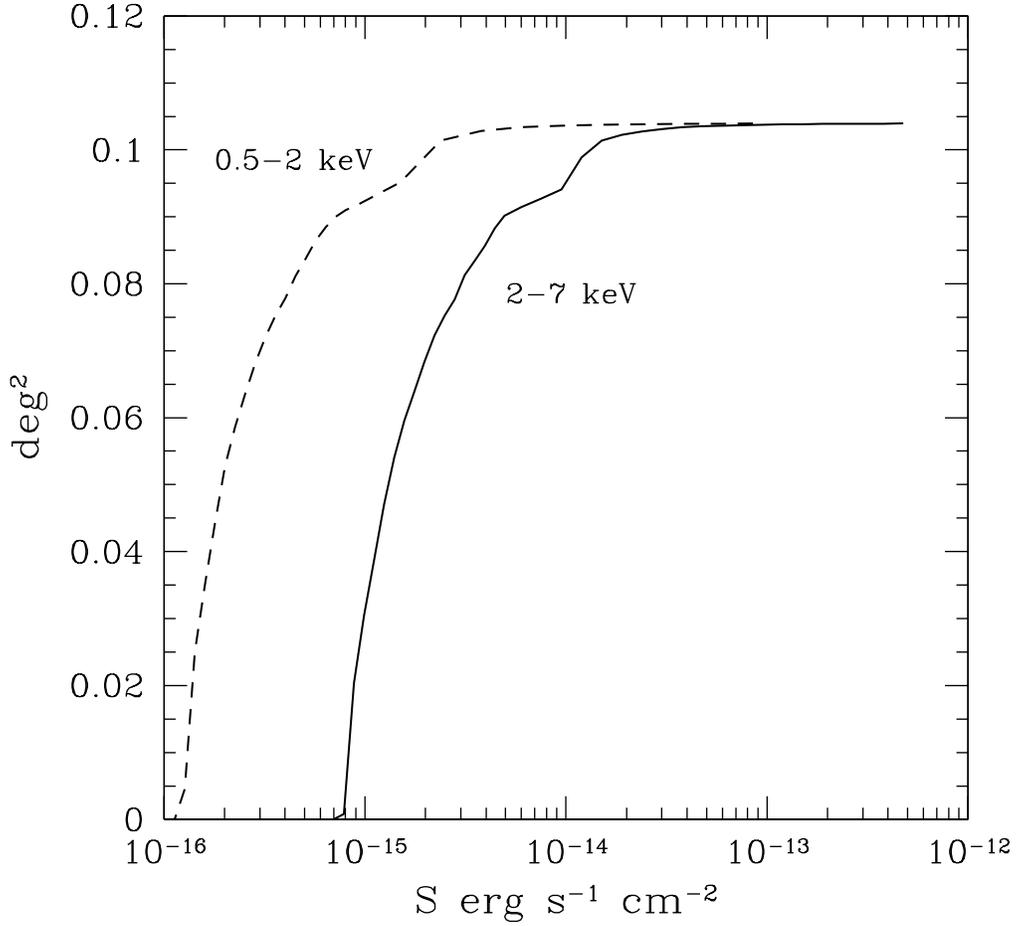,height=6.in}}
\caption{Sky coverage of the 300ks exposure of the Chandra Deep Field
South for the soft (dashed line) and the hard (solid line) band, as a
function of the soft and hard fluxes respectively.  Note that the hard
energy flux is computed in the 2--10 keV band, while the sky coverage
refers to the 2-7 keV image.  The step--like features are due to the
different roll angles in the 4 observations that produce corners with
different exposure times.
\label{fig4}}
\end{figure}

We now calculate the effective sky coverage at a given flux which is
defined as the area on the sky where a source with a given net count
rate can be detected with a $S/N > 2.1$ in the extraction region
defined in \S 2.1.  The computation includes the effect of varying
exposure, vignetting and point response function variation across the
field.  Such a procedure has been tested extensively with simulations
performed with MARX v3.0, and it has been shown to be accurate to within
a few percent (see panel d in Figure \ref{fig1}).  The sky coverage is
given in Figure \ref{fig4}.  The step--like features are due to the
large difference in the exposure across the field of view.

\begin{figure}
\centerline{\psfig{figure=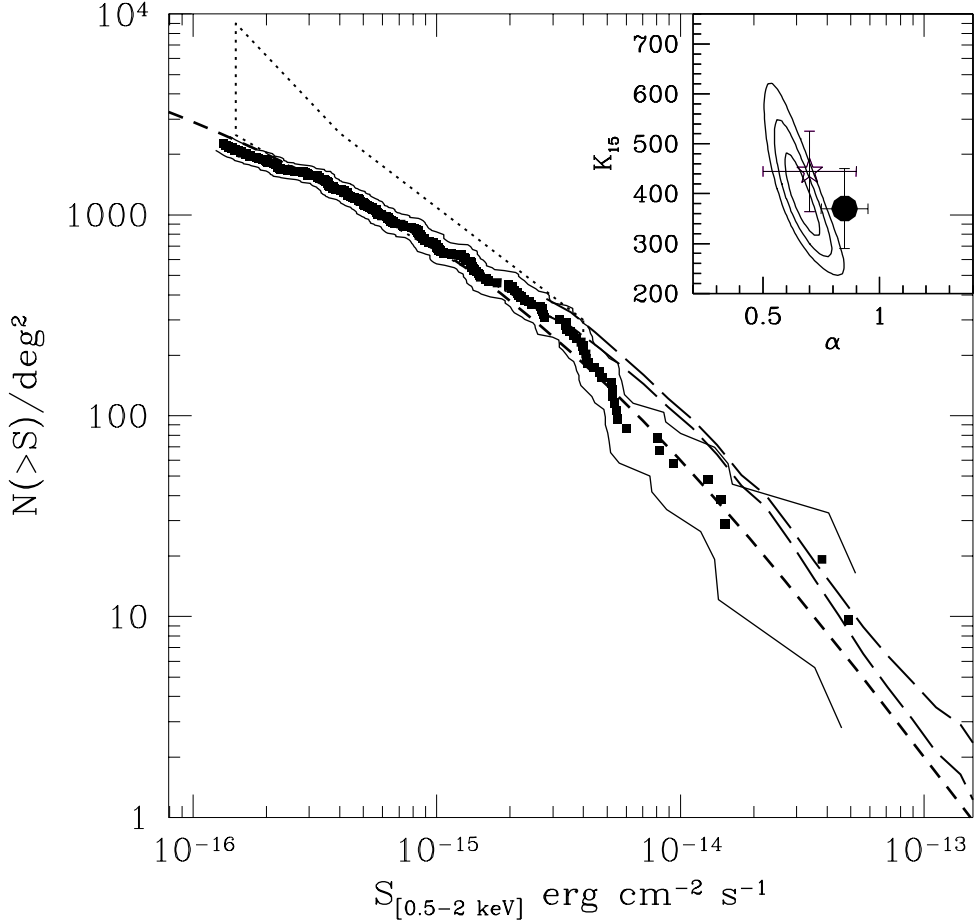,height=6.in}}
\caption{The LogN--LogS in the soft band from the Chandra Deep Field
South for an average spectral slope $\Gamma = 1.4$ (filled squares).
Dashed lines are the counts from ROSAT observation of the Lockman Hole
(Hasinger et al. 1998), and the dotted contour is the extrapolation
from the fluctuation analysis in ROSAT data (Hasinger et al. 1993).
The upper and lower solid lines indicate uncertainties due to the sum
of the Poisson noise (1 sigma) including the uncertainty in the
conversion factor (see text).  The short dashed line is from model B
in Gilli, Salvati \& Hasinger (2001), rescaled to fit the total
background measured by HEAO--1 (see text).  The insert shows the
maximum likelihood fit to the parameters in the LogN--LogS fit $
N(>S)=K_{15}(S/ {2 \times 10^{-15}})^{- \alpha}$. The contours
correspond to $1 \sigma$, $2 \sigma$ and $3 \sigma$.  The star is the
fit from Mushotzky et al. (2000) at $S=2 \times 10^{-15}$ erg s$^{-1}$
cm$^{-2}$; the error bar is their quoted 68 \% confidence level.  The
large dot is the result from the 120k exposure (for $\Gamma = 1.4$).
\label{fig5}}
\end{figure}

Using the sky--coverage we can compute the number counts in the soft
and hard bands.  In the soft band, shown in Figure \ref{fig5}, we find
the Chandra results in excellent agreement with ROSAT in the region of
overlap $S_{min} >10^{-15}$ erg s$^{-1}$ cm$^{-2}$. The Chandra data
extend the results to $10^{-16}$ erg s$^{-1}$ cm$^{-2}$.  A maximum
likelihood fit to the soft sources gives
\begin{equation} N(>S) = (400\pm 80) \Big( {{S}\over
{2\times 10^{-15}}}\Big) ^{-0.66\pm 0.06}\, \, \, \, \, \,  {\rm cgs}
\end{equation} 
where the errors are at 1 sigma.  The confidence levels at 1, 2 and 3
sigma for both slope and normalization are shown in the insert panel
of Figure \ref{fig5}.  The clear flattening with respect to the bright
counts by ROSAT ($S>10^{-13} $ erg s$^{-1}$ cm$^{-2}$) shown in Paper
I (also found by Mushotzky et al. 2000) is confirmed.  The slope shows
2$\sigma$ evidence of further flattening with respect to the value
found in Paper I.  This difference may be partially due to the number
of spurious sources and flickering pixels not removed in the previous
data reduction, causing the soft number counts at the lowest fluxes to
be overestimated by $\simeq 15$\%.  To check if the number counts are
flattening in the investigated flux range, we perform a fit with a
double power law for the differential number counts.  The fit is done
on 4 parameters: faint--end normalization and slope, bright--end slope
and flux where the break occurs.  We find that the differential counts
are well fitted by a power law which is consistent with the euclidean
slope at the bright end and with a slope $\alpha_{diff} \equiv
\alpha+1 = 1.5 \pm 0.1$ (1 sigma) at the faint end, with a break at $S
= 1.1\times 10^{-14}$ erg s$^{-1}$ cm$^{-2}$.  Thus, below
$S=10^{-15}$ erg s$^{-1}$ cm$^{-2}$, the slope of the cumulative
number count is $\alpha \simeq 0.5$.  Such a significant flattening
will be further tested on the data from the 1Ms observation.

We compared our LogN--LogS with the predictions of the AGN population
synthesis models described in Gilli, Salvati \& Hasinger (2001).  We
consider their model B, where the number ratio R between absorbed and
unabsorbed AGNs is assumed to increase with redshift from 4, the value
measured in the local Universe, to 10 at $z\sim 1.3$, where R is
unknown.  This model was found to provide a better description of the
X--ray constraints (ASCA and ROSAT number counts, luminosity function
and redshift distribution by ROSAT, see Gilli, Salvati \& Hasinger
2001) with respect to a standard model where R=4 at all redshifts.
The value of R was independent of the luminosity and therefore a large
population of obscured QSOs is included in the model.  Such a model
was originally calibrated to fit the background intensity measured by
ASCA.  In this paper, the parameters of the AGN X--ray luminosity
functions assumed in the model B are recalibrated to fit the
background intensity measured by HEAO--1.  The predictions for the
source counts (short dashed line in Figure \ref{fig5}) are in very
good agreement with the Chandra LogN-\-LogS at all fluxes.

For the total contribution to the soft X--ray background, we refer to
the 1--2 keV band, following Hasinger et al. (1993; 1998) and
Mushotzky et al. (2000).  In this energy band the comparison with the
unresolved soft X--ray background is more straightforward than in the
0.5--2 keV band, since for energies $<1$ keV the Galactic contribution
may be significant.  The unresolved value in the 1--2 keV band
measured by ROSAT is $4.4\times 10^{-12}$ erg s$^{-1}$ cm$^{-2}$,
while the value found by ASCA is $3.7\times 10^{-12}$ erg s$^{-1}$
cm$^{-2}$ (see Hasinger et al. 1998 and references therein).  We find
a contribution of $\simeq 5.6 \times 10^{-13} $ erg s$^{-1}$ cm$^{-2}$
deg$^{-2}$ for fluxes lower than $10^{-15}$ erg s$^{-1}$ cm$^{-2}$,
corresponding to $\simeq 13-15$\% of the unresolved flux.  If this
value is summed to the contribution at higher fluxes ($3.02 \times
10^{-12}$ erg s$^{-1}$ cm$^{-2}$ deg$^{-2}$, see Hasinger et
al. 1998), we end up with a total contribution of $\simeq 3.57 \times
10^{-12} $ erg s$^{-1}$ cm$^{-2}$ deg$^{-2}$ for fluxes larger than
$0.6 \times 10^{-16}$ erg s$^{-1}$ cm$^{-2}$ (which is our flux limit
in the 1--2 keV band), corresponding to 81\% of the unresolved value
as measured by ROSAT.  If we assume the unresolved value measured by
ASCA (which must be considered as a lower limit) the resolved
contribution amounts to 96\%.

\begin{figure}
\centerline{\psfig{figure=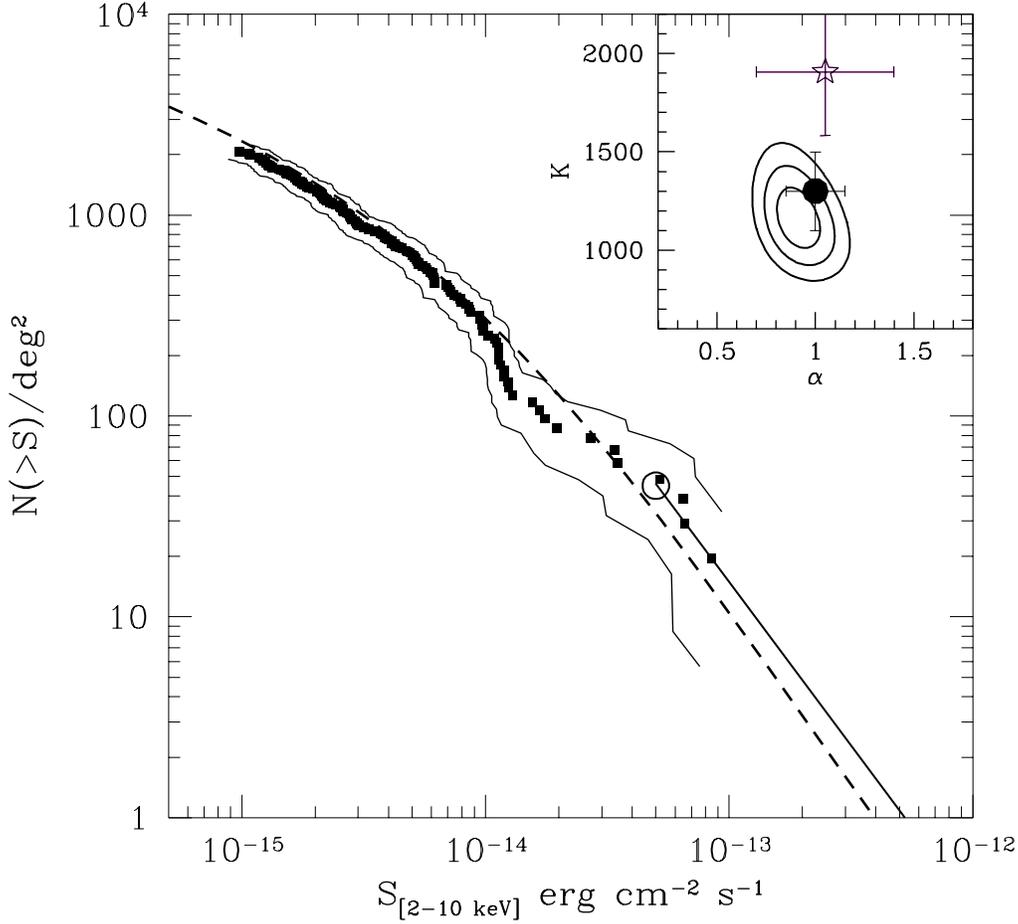,height=6.in}}
\caption{The LogN--LogS in the hard band from the Chandra Deep Field
South for an average spectral slope $\Gamma = 1.4$, symbols as in
Figure \ref{fig5}.  The open circle at high fluxes is from ASCA and
Beppo SAX (Giommi, Perri \& Fiore 2000, Ueda et al. 1999) and the
continuous line is the fit to the counts from ASCA in the range
$10^{-12}-10^{-13}$ erg cm$^{-2}$ s$^{-1}$ (Della Ceca et al. 1999b).
The upper and lower solid lines indicate uncertainties due to the sum
of the Poisson noise (1 $\sigma$) including the uncertainty in the
conversion factor (see text).  The insert shows the maximum likelihood
fit to the parameters in the LogN--LogS fit $N(>S)=K_{15}(S/{2 \times
10^{-15}})^{-\alpha}$.  The thick contours correspond to $1\sigma$,
$2\sigma$ and $3\sigma$.  The star is the fit from Mushotzky et
al. (2000) at $S=2 \times 10^{-15}$ erg s$^{-1}$ cm$^{-2}$; the error
bar is their quoted 68 \% confidence level.  The large dot is the
result from the 120k exposure (for $\Gamma = 1.4$).  The dashed line
is the same model of Figure \ref{fig5}.
\label{fig6}}
\end{figure}

We show the LogN--LogS distribution for sources in the hard band in
Figure \ref{fig6}, down to a flux limit of $10^{-15}$ erg s$^{-1}$
cm$^{-2}$.  The hard counts are normalized at the bright end with the
ASCA data by Della Ceca et al. (1999b).  A maximum likelihood fit to
the hard sources data is:
\begin{equation} N(>S) = (1150 \pm 150) \Big( {{S}\over {2\times 10^{-15}}}\Big)
^{-0.92\pm 0.12}\, \, \, \, \, \,   {\rm cgs},  
\end{equation}
where the errors are at 1 sigma.  We used a conversion factor for an
average power law $\Gamma=1.4$.  The confidence contours at 1, 2 and 3
sigma for the slope and the normalization are shown in the upper right
corner of Figure \ref{fig6}.  The hard number counts confirm the
finding of Paper I (large filled dot) of a substantial flattening with
respect to the slope at the bright end ($\alpha = 1.67\pm 0.09$, Della
Ceca et al. 1999b).  We repeat the fit with a double power law in the
differential counts, as in the case of the soft counts.  For the hard
counts too, we find evidence for further flattening: the differential
counts are well fitted by a power law which is consistent with the
euclidean slope at the bright end and with a slope $\alpha_{diff} =
1.6 \pm 0.2$ (1 sigma) at the faint end, with a break at $S = 9\times
10^{-15}$ erg s$^{-1}$ cm$^{-2}$.  To confirm such change in the slope
below $S=10^{-14}$ erg s$^{-1}$ cm$^{-2}$ we need to extend the number
counts to a lower flux limit.

The normalization computed at $2\times 10^{-15}$ erg s$^{-1}$
cm$^{-2}$ is consistent with the best fit in Paper I.  Thus, we still
find a difference larger than 3 sigma from the results of Mushotzky et
al. (2000, large star).  We stress that here we used an average
spectral slope ($\Gamma = 1.4$) flatter than in PaperI (where we used
$\Gamma = 1.7$).  We chose the new value to be in agreement with the
average spectral shape of our new combined sample. The short dashed
line in Figure \ref{fig6} is the same model as in Figure \ref{fig5}.

Despite the assumption of a flatter spectral slope, the difference in
normalization with the hard number counts of Mushotzky et al. (2000)
is still present.  Our hard number counts are confirmed also by the
results from the Lynx field, independently analysed by some of us, and
from the XMM observation of the Lockman Hole (Hasinger et al. 2001a).
The disagreement between us and Mushotzky et al. (2000) is
statistically significant and acquires particular relevance since we
are about to resolve completely the hard background.  The causes of
discrepancy may be due to calibration problems\footnote{A 15\%
underestimation of the effective area of the BI chips has been
recently discovered, but only below 1.2 keV, see
http://asc.harvard.edu/ciao/caveats/effacal.html)}.  A possible
physical reason is cosmic variance.  In this case the larger area of
the CDFS should give a more accurate measure, if compared with the
smaller area (1/4) covered by the pointing of Mushotzky et al. (2000).

\begin{figure}
\centerline{\psfig{figure=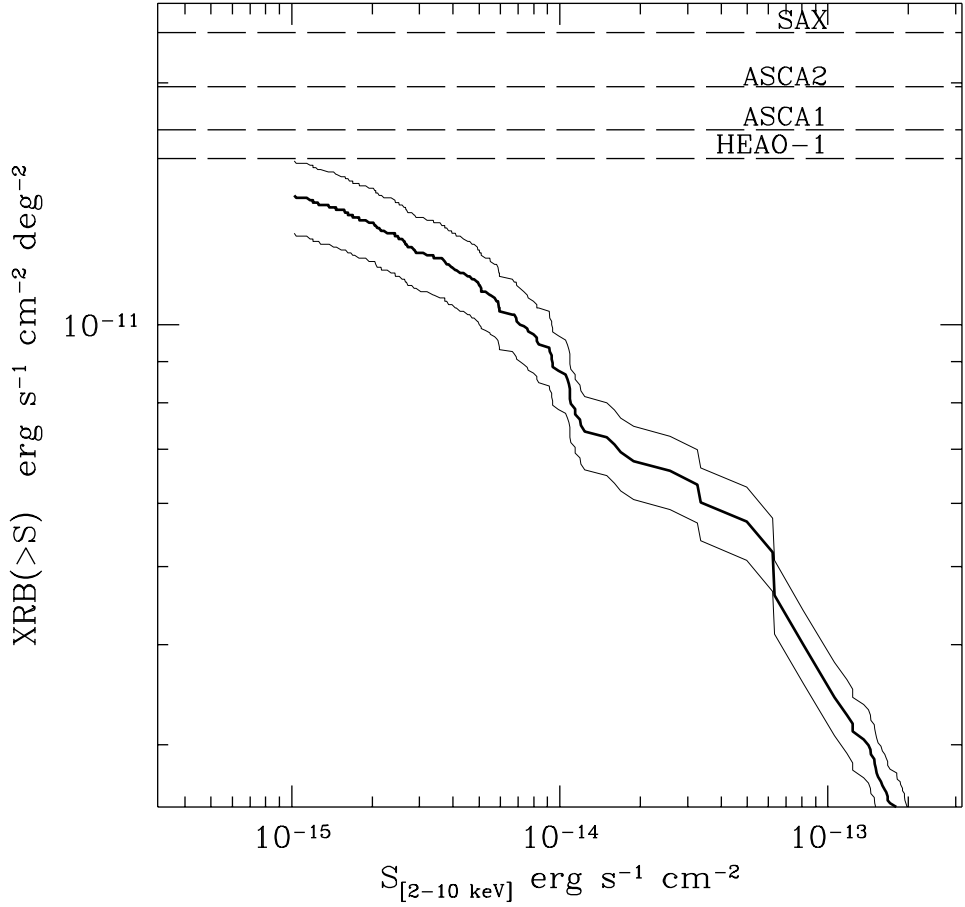,height=6.in}}
\caption{The contribution to the hard X--ray flux density as a
function of the flux of the resolved sources.  The total resolved
contribution is computed from the CDFS sample plus the bright sample
from ASCA at fluxes larger than $\simeq 10^{-13}$ erg s$^{-1}$
cm$^{-2}$ (Della Ceca et al. 1999b).  The upper dashed lines refer to
previous measures of the hard X--ray background; from bottom to top:
Marshall et al. (1980, HEAO-1), Ueda et al. (1999, ASCA1), Ishisaki
(1999, ASCA2), Vecchi et al. (1999, BeppoSAX).
\label{fig7}}
\end{figure}

The integrated contribution of all the sources within the flux range
$10^{-13}$ erg s$^{-1}$ cm$^{-2}$ to $ 10^{-15}$ erg s$^{-1}$
cm$^{-2}$ in the 2--10 keV band is $(1.15 \pm 0.15) \times 10^{-11}$
erg s$^{-1}$ cm$^{-2}$ deg$^{-2}$ for $\Gamma=1.4$.  After including
the bright end seen by ASCA for $S> 10^{-13}$ erg s$^{-1}$ cm$^{-2}$
(Della Ceca et al. 1999b), the total resolved hard X--ray background
amounts to $(1.46 \pm 0.20) \times 10^{-11}$ erg s$^{-1}$ cm$^{-2}$
deg$^{-2}$.  The inclusion of the ASCA data at bright fluxes minimizes
the effect of cosmic variance.  In Figure \ref{fig7} we show the total
contribution computed from the CDFS plus ASCA sample.  The value at
the limiting flux $S=10^{-15}$ erg s$^{-1}$ cm$^{-2}$ is still lower
than the value of the total (unresolved) X--ray background $1.6 \times
10^{-11}$ erg s$^{-1}$ cm$^{-2}$ deg$^{-2}$ from UHURU and HEAO-1
(Marshall et al. 1980).  More recent values of the 2--10 keV
integrated flux from the BeppoSAX and ASCA surveys (e.g., Vecchi et
al. 1999; Ishisaki et al. 1999 and Gendreau et al. 1995) appear higher
by 20-40\%.  These latest values are closer to the old Wisconsin
measurements which were about 30\% higher than the HEAO-1 value. The
difference of 30\% in flux has never been resolved up to the current
day (McCammon \& Sanders 1990).

\begin{figure}
\centerline{\psfig{figure=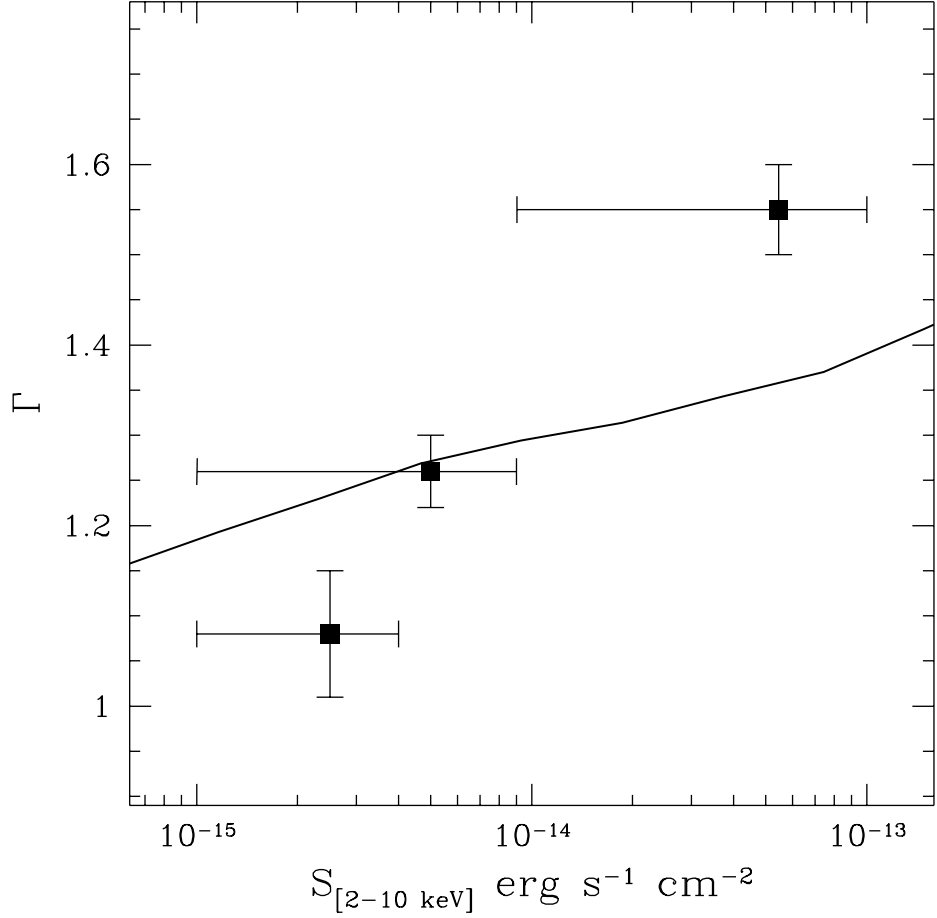,height=6.in}}
\caption{The average power law index of the stacked spectra of the
bright, faint and very faint subsamples of the sources detected in the
hard band, as defined in the text.  Errors on $\Gamma$ refer to the
90\% confidence level.  The local absorption has been fixed to the
Galactic value $N_H = 8\times 10^{19}$ cm$^{-2}$. The line is derived
from the same model of Figure \ref{fig5} and \ref{fig6}.
\label{fig8}}
\end{figure}

Before concluding that a relevant fraction (10-40\%)of the XRB has
still to be resolved, we perform a detailed analysis of the hard
sample, in order to understand if the resolved contribution may be
underestimated due to a wrongly assumed spectral shape.  To have a
better understanding of how the spectrum of the X--ray background is
built up at different fluxes, we divide the hard band sample
into a bright subsample ($9 \times 10^{-15} < S < 10^{-13}$ erg
s$^{-1}$ cm$^{-2}$), a faint subsample ($10^{-15} < S < 9\times
10^{-15}$ erg s$^{-1}$ cm$^{-2}$) and a very faint subsample
($10^{-15} < S < 4 \times 10^{-15}$ erg s$^{-1}$ cm$^{-2}$) in terms
of the hard fluxes.  The sources were selected only in the area with
300ks of exposure, and the stacked spectra contain about 3300, 3700
and 1550 net counts respectively.  We fitted the energy range from 1
keV to 10 keV with an absorbed power law and fixed the local
absorption to the Galactic value $ N_H = 8 \times 10^{19}$ cm$^{-2}$.
The best--fit photon index is $\Gamma = 1.55 \pm 0.03$ for the bright
sample, $\Gamma = 1.26 \pm 0.04$ for the faint sample, and $\Gamma =
1.08 \pm 0.07$ for the very faint sample (1 sigma errors).  The trend,
shown in Figure \ref{fig8}, is representative of the progressive
hardening of the sources at lower fluxes. The line in Figure
\ref{fig8} is the prediction from the model used in Figures \ref{fig5}
and \ref{fig6}.  Here we notice that at the very bright end our sample
is dominated by few bright sources with a spectrum steeper than the
rest of the sample, typical of the TypeI AGNs detected by ASCA at
$S>10^{-13}$ erg s$^{-1}$ cm$^{-2}$ (which make about 30\% of the hard
background and have $\Gamma \simeq 1.8$).  This may have biased the
average spectral slope of our total sample towards slightly higher
values.  For example, the average spectral slope in the Lynx field,
independently analyzed by some of us, in the same range of fluxes is
$\Gamma = 1.36 \pm 0.02$ (1 sigma error) which is marginally flatter
than $\Gamma = 1.44 \pm 0.03$ found in the CDFS.

We checked that using a varying $\Gamma$, as shown in Figure
\ref{fig8}, does not alter the resulting logN--logS by more than a 5\%
increase in the normalization.  The total contribution to the X-ray
background changes even less, since the lower $\Gamma$ at low fluxes
is compensated by the higher $\Gamma$ at high fluxes.  We conclude
that using an average $\Gamma = 1.4$ to derive the hard fluxes from
the net counts is justified. An interesting consequence of this
analysis is that a fraction of at least $10$\% of the unresolved hard
XRB is still to be resolved at fluxes lower than $10^{-15}$ erg
s$^{-1}$ cm$^{-2}$.


\section {PROPERTIES OF SOURCES}

The spectroscopic optical identification process of the X--ray sources
described in this paper is still ongoing. Spectra have been taken with
the FORS1 instrument on the VLT UT-1 in multislit mode. From about 130
spectra taken, we obtained a total of 99 redshifts, of which we regard
86 as secure.

The redshifts are distributed between $z=0.1$ and $z = 3.7$ with 50
sources at $z \leq 1$ and 36 at $z > 1$.  In Figure \ref{fig9} we plot
the distribution as a function of redshift and hardness ratio.  We
notice a striking difference between the distribution of hard and soft
sources.  The large majority of the sources with $HR>0$ appear at
$z\leq 1.6$, most of them at $z<1$.  Moreover, most of the sources
with $HR>0$ are identified as TypeII AGNs from the presence of narrow
lines.  TypeI AGNs instead, show an average $HR\simeq -0.5$ with a
relatively narrow dispersion, and they are found in a broader range of
redshift.

Only three sources have $HR\geq 0$ and $z>2$, including an obscured
QSO at $z=3.7$ with $HR\simeq 0$.  This source is marked with an
asterisk in Figure \ref{fig9} and is studied in detail in Norman et
al. (2001).  The average hardness ratio of all sources at $z> 1.6$ is
$HR\simeq -0.5$.  This result, if confirmed by the complete analysis
of the sample, would indicate that the hardest sources contributing to
the 2--10 keV background are typically TypeII objects at $z \leq 1$,
according to what has been found by Barger et al. (2001).  Conversely,
the most distant sources ($z>2$) have an average $HR \simeq -0.5$ and
are prevalently TypeI objects.  The dashed lines in Figure \ref{fig9}
show how the observed hardness ratio changes with redshift for a given
intrinsic absorbing $N_H$ assuming a photon index $\Gamma =1.7$.  We
note that $HR$ is not a good tracer of intrinsic absorption especially
at high $z$.

We can investigate in greater detail the nature of the hard sources.
We selected a subsample of source with $HR >0$ (corresponding to an
effective $\Gamma < 0.5$ for a Galactic $N_H$).  All the selected
sources are included in the region with 300ks of exposure and have a
hard flux lower than $2 \times 10^{-14}$ erg s$^{-1}$ cm$^{-2}$.  The
sources selected in this way contribute about 20-30\% of the total
X--ray background.  A fit with an absorbed power law gives $ \Gamma =
1.23\pm 0.13$ and $N_H = (1.3\pm 0.2)\times 10^{22}$ cm$^{-2}$ (1
sigma errors). Using an average redshift of $\langle z \rangle \sim
0.7$ as representative of the sources in the subsample, we find an
average intrinsic absorption corresponding to $N_H \simeq (4\pm
0.4)\times 10^{22}$ cm$^{-2}$ with a similar $\Gamma$.  This confirms
the presence of strong intrinsic absorption for the hardest sources
detected at low fluxes.

Luminosities are computed in the rest--frame soft band after assuming
an average power law spectrum with $\Gamma = 1.4$, in a critical
universe with $H_0 = 50$ km/s/Mpc.  The distribution of luminosities
emitted in the soft band with redshift is shown in Figure \ref{fig10}.
This plot shows that in 300ks we have achieved a greater sensitivity
than all previous surveys.  The upper solid line corresponds to a
sensitivity limit in flux of $2\times 10^{-16}$ erg s$^{-1}$
cm$^{-2}$, that was the flux limit of the 120k exposure (see Paper I).
Most of the sources with $HR>0$ appear to fall in the $L_{hard} =
10^{42} - 10^{44}$ erg s$^{-1}$ range and are identified with TypeII
absorbed AGNs.  Data obtained with the HST observations of a small
fraction of our field (Schreier et al. 2001) will address in detail
the morphological properties of the optical counterparts.

Softer sources with $HR<0$ appear to span the range from $10^{40}$ to
$10^{45}$ erg s$^{-1}$.  At luminosities larger than $10^{42}$ erg
s$^{-1}$ these sources are mostly identified with TypeI AGN.  At the
low luminosity end, several sources are detected only in the soft band
($HR=-1$).  Most of these objects are identified with galaxies at
redshift less than $0.6$ with soft luminosities restricted to a range
between $L_x =10^{40}$ and $L_x = 10^{42}$ erg s$^{-1}$.  This
population of relatively nearby galaxies with soft X--ray spectra has
been found also by Hornschemeier et al. (2001) in the Chandra Deep
Field North.  

In the plot $S$ versus $R$, where S is the X--ray flux in the soft
band (see Figure 7 of Paper I) this subsample appears to populate the
low $S/R$ region at fluxes $S < 10^{-15}$ erg s$^{-1}$ cm$^{-2}$.  We
can use a criterium defined in the $S$--$R$ plane to isolate this
soft, low luminosity population of sources.  If we define a subsample
requiring $S/S_{opt}<0.1$, $S_{tot} < 3\times 10^{-15}$ erg s$^{-1}$
cm$^{-2}$, and $HR < -0.5$, we find 18 sources.  Among these 18
sources, 10 have redshift, with an average value $\langle z \rangle =
0.25$.  The very high redshift sources with $HR = -1$ (see Figure
\ref{fig9}) have $S/S_{opt} \simeq 1$ and are therefore excluded from
this subsample.  By visual inspection from the images in the $R$ and
$B$ bands, at least 10 of these 18 sources appear clearly to be
galaxies.  If we fit their stacked spectrum with a power law, we find
consistency with Galactic absorption, and a spectral slope $\Gamma =
2.0\pm 0.2$ (90 \% c.l.).  If we repeat the fit with a thermal
spectrum, we find a temperature of $ 3.5 \pm 1.5$ keV, which is too
high to be produced by hot gas in the potential wells of galaxies.
Such an apparent temperature may be due to the contribution of low
luminosity nuclear activity or a population of X--ray binaries.  Both
fits have a $\chi_\nu^2\simeq 1$.  A better definition of the
subsample will be available when the spectroscopic optical
identification is completed.

\section {CONCLUSIONS}

We detected X--ray sources down to a flux limit of $10^{-16}$ erg
s$^{-1}$ cm$^{-2}$ in the 0.5--2 keV soft band and $10^{-15}$ erg
s$^{-1}$ cm$^{-2}$ in the 2--10 keV hard band, improving the
sensitivity by a factor of 2 with respect to Paper I.  The LogN--LogS
relationship of the hard sample confirmed the previous findings
obtained from the 120 ksec exposure (Paper I).  The average slope
turns out to be $\alpha = 0.92 \pm 0.12$, which is flatter with
respect to the quasi--euclidean slope found by ASCA at the bright end
($S>10^{-13}$ erg s$^{-1}$ cm$^{-2}$).  However, adopting a fit with
double power law in the whole flux range, we find indication for
flattening down to $\alpha = 0.6 \pm 0.2$ below $S\simeq 10^{-14}$ erg
s$^{-1}$ cm$^{-2}$.

After the inclusion of the ASCA sources at the bright end, the total
contribution to the resolved hard X--ray background down to our flux
limit of $10^{-15}$ erg s$^{-1}$ cm$^{-2}$ now amounts to $(1.45 \pm
0.15)\times 10^{-11}$ erg cm$^{-2}$ s$^{-1}$ deg$^{-2}$.  This result
is robust to variations of the average spectral slope in the hard
sample, which can be as low as $\Gamma = 1.1$ for the faintest
sources. The value for the unresolved X--ray background ranges from
$1.6$ to $2.4 \times 10^{-11}$ erg s$^{-1}$ cm$^{-2}$, so a fraction
between 60\% and 90\% of the hard XRB has now been resolved.

We confirm our previous finding on the average spectrum of the sources
of the total sample, which can be well approximated with a power law
with index $\Gamma=1.4$.  We also find a progressive hardening of the
sources at lower fluxes, both for the soft and the hard samples.  In
particular, we divided the hard sample into three subsamples: bright
($9 \times 10^{-15} < S < 10^{-13}$ erg s$^{-1}$ cm$^{-2}$), faint
($10^{-15} < S < 9\times 10^{-15}$ erg s$^{-1}$ cm$^{-2}$) and very
faint ($10^{-15} < S < 4 \times 10^{-15}$ erg s$^{-1}$ cm$^{-2}$).
The average power law is $\Gamma = 1.55 \pm 0.03$ for the bright
subsample, $\Gamma = 1.26 \pm 0.04$ for the faint subsample, $\Gamma =
1.08 \pm 0.07$ for the very faint subsample.  The progressive
hardening at faint fluxes is probably due to a stronger intrinsic
absorption.

In the soft band the counts show a slope of $\alpha = 0.66\pm 0.06$ (1
sigma error), clearly flatter than the slope at the bright end
measured by ROSAT.  A double power law fit indicates a progressive
flattening at the faint end, with a best fit with $\alpha = 0.5 \pm
0.1$ for fluxes below $10^{-15}$ erg s$^{-1}$ cm$^{-2}$.  The soft
X--ray background in the energy range 1--2 keV is now resolved at the
level of 81--96 \% depending on the assumed value for the unresolved
soft XRB.

The soft and the hard LogN--logS must ultimately approach each other.
In fact, in the unified AGN explanation of the XRB we should, with
increasing sensitivity, see the same objects in both bands.  However,
if the remaining hard population is constituted by sources that are
even more strongly absorbed, there will be a significant fraction of
hard sources still undetected in the soft band also at $S \simeq
10^{-17}$ erg s$^{-1}$ cm$^{-2}$.  Following the model B in Gilli,
Salvati \& Hasinger (2001), we found that the soft and hard LogN--LogS
predict the same source density at $S\sim10^{-19}-10^{-20}$ erg
s$^{-1}$ cm$^{-2}$.

\begin{figure}
\centerline{\psfig{figure=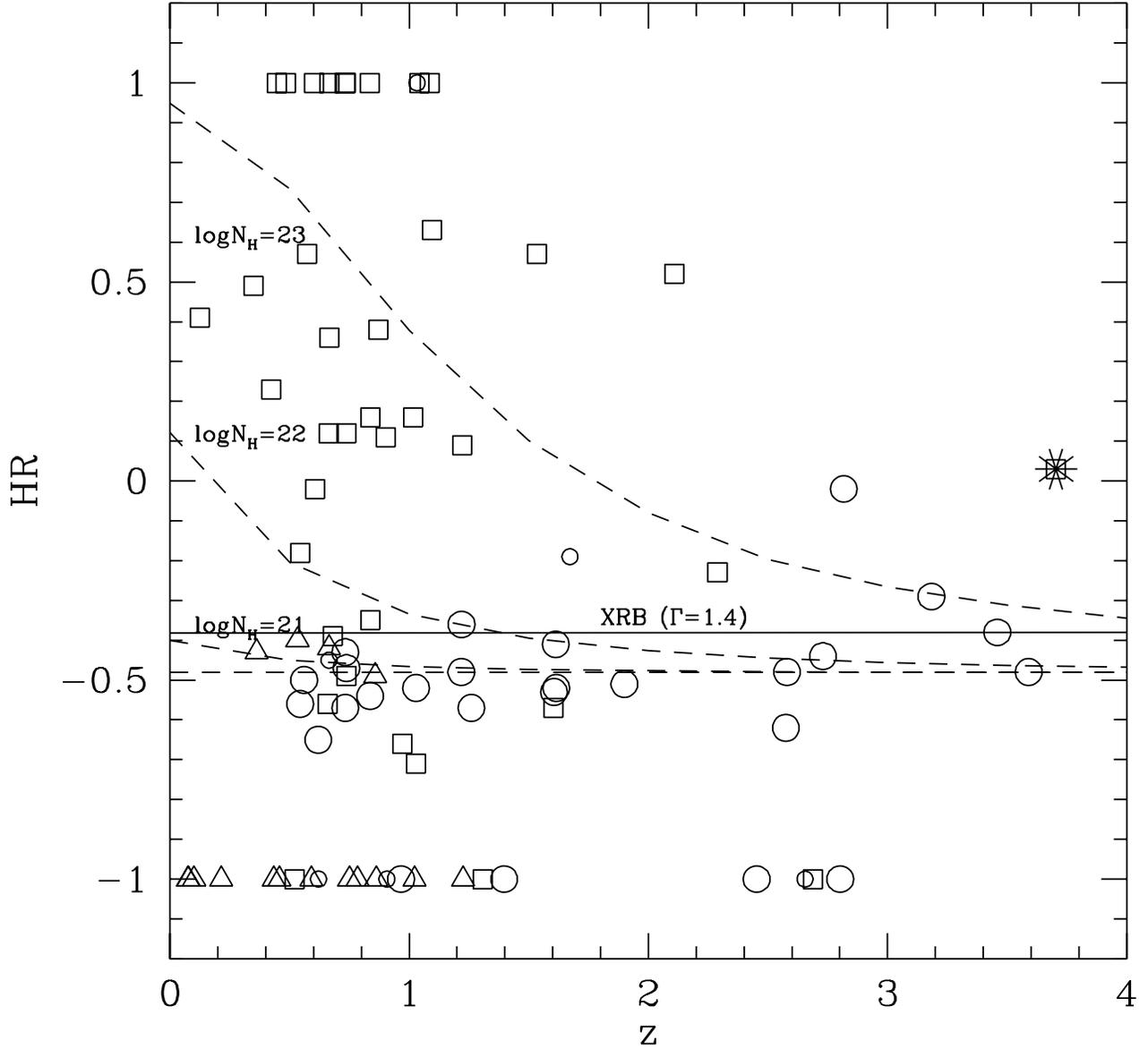}}
\caption{Hardness ratio as a function of redshift for the 86 sources
with optical spectra in our sample.  TypeI AGNs are marked with large
open circles; TypeII with open squares; galaxies with triangles.
Objects with uncertain type identification are marked with small
circles.  The dashed lines show how the hardness ratio change with $z$
for a given intrinsic absorbing column assuming a photon index $\Gamma
=1.7$.  The horizontal solid line is the average hardness ratio of the
total XRB ($\Gamma = 1.4$).  The source marked with a star is the
absorbed QSO discussed in Norman et al. (2001).
\label{fig9}}
\end{figure}

\begin{figure}
\centerline{\psfig{figure=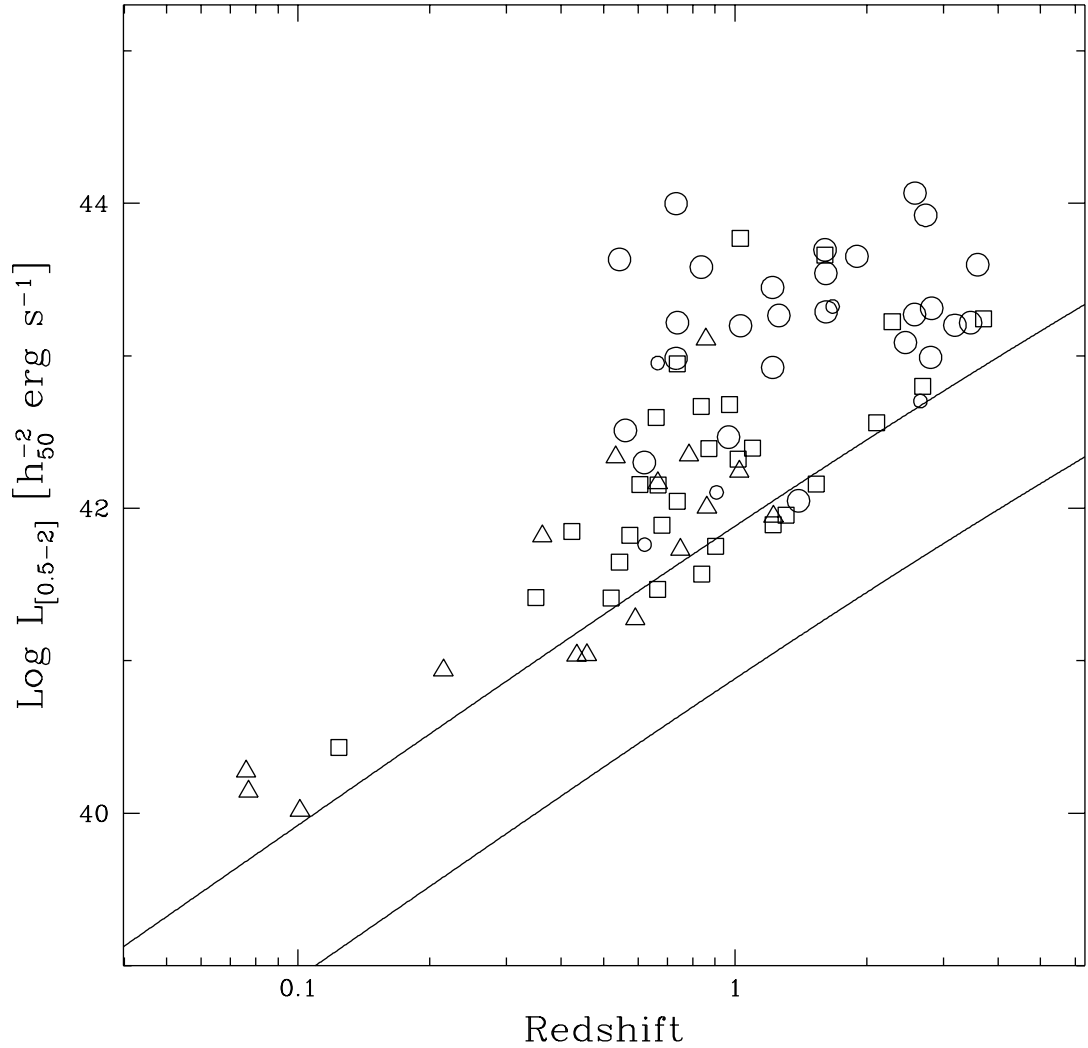,height=6.in}}
\caption{$L_X$ in the soft band (rest frame) versus redshift for the
Chandra sources.  Symbols as in Figure \ref{fig9}.  The two solid
lines correspond to a flux limit of $ 2\times 10^{-16}$ erg cm$^{-2}$
s$^{-1}$ (upper line, the flux limit of the previous 120 ks exposure)
and to $ 2\times 10^{-17}$ erg cm$^{-2}$ s$^{-1}$ (lower line),
assuming a spectral slope $\Gamma =1.4$.  A critical universe with
$H_0 = 50$ km/s/Mpc has been assumed.
\label{fig10}}
\end{figure}

We have also used measured redshifts for 86 sources.  The spectra will
be presented in the catalog paper of the 1Ms exposure (Giacconi et
al, in preparation).  We find that the hard sources with $HR >0$ are
typically found at $z<1$, with intrinsic luminosities between
$10^{42}$ and $10^{44}$ erg s$^{-1}$ in the soft band.  Between the
objects producing the 2--10 keV X--ray background, the ones with $HR
>0$ (producing between 20--30\% of the total XRB) appear relatively
nearby (redshifts $z<1$) and are most easily identified with Seyfert
II galaxies (see also Barger et al. 2001).  We note that a bias could
be present against obscured AGNs, since our spectroscopically
identified sample is not complete and obscured AGNs are expected to
have on average fainter magnitudes than the unobscured ones.  One of
the important results of Chandra is the detection of obscured QSOs.  A
first clear example of hard (i.e. likely obscured) X--ray sources at
high redshift with X--ray luminosities above $3\times 10^{44}$ erg
s$^{-1}$ have been already found (see Figure \ref{fig9}).  This source
is extensively described in Norman et al. (2001).  When the optical
identification program will be completed, it will be possible to test
whether the population of aborbed AGN is as large as assumed in the
AGN synthesis models of the XRB (Madau, Ghisellini \& Fabian 1993;
Comastri et al. 1995; Gilli et al. 2001, Comastri et al. 2001).

The observation at $S < 10^{-15}$ erg s$^{-1}$ cm$^{-2}$ of very soft
($HR<-0.5$) objects with $L_X$ between $ 10^{40}$ and $10^{42}$
erg s$^{-1}$ up to $z\simeq 0.3$ opens up a new window to the study
of the evolution of galaxies at moderate $z$.  A large number of galaxies
could increase the number counts at low fluxes without contributing
significantly to the XRB.

\acknowledgements

G. Hasinger and G. Szokoly acknowledge support under DLR grant 50 OR
9908 O.  R. Giacconi and C. Norman gratefully acknowledge support
under NASA grant NAG-8-1527 and NAG-8-1133.  We thank Richard Bouwens
for providing the modified version of Sextractor used for detection
and Roberto Della Ceca, Ken Kellerman and Peter Shaver for useful
comments.  We thank the anonymous refere for a constructive and
detailed report.

\newpage

\end{document}